\documentclass[twocolumn]{revtex4}

\def\be{\begin{equation}}
\def\ee{\end{equation}}
\def\la{\label}
\def\bea{\begin{eqnarray}}
\def\eea{\end{eqnarray}}

\def\ci{\cite}
\def\la{\label}
\def\bib{\bibitem}
\def\lesssim{{_ <\atop{^\sim}}}
\def\lm{\lambda}

\def\le{\left}
\def\ri{\right}

\def\Ompo{\Omega_{\phi o}}

\def\s8{\sigma_8}

\def\fr{\frac}
\def\pp{\partial}
\def\pu{\pp_\mu}
\def\pU{\pp^\mu}

\def\vp{\varphi}

\def\Op{\Omega_\phi}
\def\Opo{\Omega_{\phi o}}

\def\Odmo{\Omega_{dm o}}

\def\rdmo{\rho_{dm o}}
\def\rdm{\rho_{dm}}

\def\rp{\rho_\phi}
\def\rpd{\rho_{\phi d}}

\def\rpo{\rho_{\phi o}}

\def\Ompo{\Omega_{\phi o}}

\def\G{\Gamma}

\def\lin{L_{int}}
\def\vin{V_{int}}

\def\vip{V_{int,\vp}}

\def\mmp{m^2_{\phi }}
\def\mmpo{m^2_{\phi o}}

\def\lin{L_{int}}
\def\mp{m_{\phi}}
\def\mpo{m_{o}}

\usepackage{graphicx}

\begin{document}

\title{Inflation-Dark Matter unified through Quantum Generation}
\author{A. de la Macorra and F. Briscese }
\affiliation{Instituto de F\'{\i}sica, Universidad Nacional
Autonoma de Mexico, Apdo. Postal 20-364,
01000 M\'exico D.F., M\'exico\\
Part of the Collaboration Instituto Avanzado de Cosmologia}

\begin{abstract}

We  unify inflation and dark matter via a single
scalar field $\phi$. One of the main difficulties for
this unification is that between inflation and dark matter
one needs a successful reheating process and a long lasting
period of radiation. Therefore the amount of energy density
in the inflaton-dark matter field $\phi$  must be fine tune after reheating
to give dark matter.
Here we show an alternative scheme in which
the  inflaton decays completely, disappearing
entirely from the spectrum. However, at low
energies, before matter-radiation equality,
the same interaction term that leads to the inflaton
decay, regenerates $\phi$.
An essential feature is that the transition between the
intermediate radiation dominated to the dark matter phase is related
to a quantum generation of the scalar field $\phi$ instead to
purely classical dynamics. Thanks to this quantum transition
the inflation-dark matter unification can take place naturally  without
fine tuning.
The unification scheme presented here has three parameters, the
mass of the dark matter particle $m_o$, the inflation parameter $\lm$ and
the coupling $g$ for the inflaton interaction. Phenomenology  fixes the value
for $\lm$ and gives a constraint  between $g$ and $m_o$, leaving
only the mass of the dark matter particle $m_o$ as a free parameter.
These same three parameters $\lm,m_o,g$  are present in models
with inflation and a dark matter wimp particle but without unification. Therefore
our unification scheme does not increase the number
of parameters and it accomplishes the desired unification between
inflaton and dark matter for free.

\end{abstract}


\maketitle

\section{Introduction}

Inflation has now become part of the standard model of
the cosmology  \cite{inflationarycosmology}. With new data coming soon,
and in particular with
the Planck satellite mission, the different inflationary models
will need to pass a strong test. Inflation set up the initial
perturbations from which gravity forms the large scale structures LSS
in the universe.
At the same time, a huge amount of evidence is gathering in
structure formation via the large galaxy maps that are currently
under way \ci{sdss}. These surveys have detected not long ago
the Baryon Acoustic Oscillation  BAO \ci{sdss}, which have the same origin
as the CMB \ci{wmap} but are measured via galaxy distribution. The formation  structure requires
the existence of  dark matter and therefore the nature of inflation and of dark matter are then
essential building blocks to understand the observations.

Inflation is  associated with a scalar field, the "inflaton", and
the energy scale at which inflation occurs is typically of the
order of $E_{I}= 10^{16} GeV$ \cite{inflationarycosmology}  but it
is possible to have consistent inflationary models with $E_I$ as
low as  $O(100) MeV$\cite{lowinflation}.  On the other hand
dark matter is described by an energy density which redshifts
as $\rdm\sim a(t)^{-3}$ and is described by particles where
its mass $m\gg T$ with $T$ its temperature. These particles
can be either fermions or scalar fields. In the case of scalar
fields the scalar potential must be $V(\phi)=m_o^2\phi^2/2$
and independently on the value of the its mass  the classical
equation of motion  ensures that $\rp\sim V \sim a(t)^{-3}$.

Here, since we want to unify inflation with dark matter we will
assume that DM is made out of the same field, a scalar field $\phi$.
\ci{inf-dm}.
A scalar field can easily give inflation and DM if
the potential is flat at high  energies and
at low energies the potential approaches the limit $V(\phi)=m_o^2\phi^2/2$.
However, most of the time our universe was  dominated by radiation. Therefore, any
realistic model must not only explain the two stages of inflation and dark matter
but must also allow for a long period of radiation domination.
Typically the inflaton decays while it oscillates around
the minimum of its quadratic potential \ci{inflationarycosmology}.
If the inflaton decay is not complete
then the remaining energy density of the inflaton after reheating
must be fine tuned to give the correct amount dark matter.

In order to avoid a fine tuning problem we follow the quantum generation
work presented in \ci{fabios}. In \ci{fabios} the quantum re-generation
process was used to unify inflation with dark energy but here we
use the same idea to unify inflation with dark matter.
To reheat the universe  we couple $\phi$
to a relativistic field $\vp$ via an interaction term $\lin$. This field $\vp$
may be a standard  model "SM" particle,
as for example neutrinos, but it could also be an extra relativistic particle not contained in the SM.
An extra relativistic degree of freedom is fine with the cosmological data \ci{rel.d}.
After inflation the $\phi$  decays into this field $\vp$ at $E_{Dinf}$ and  to reheat the universe with
"SM" particles we couple them to $\vp$  at high energies, e.g. the same
energy as the decay of $\phi$. Thermal equilibrium "TE" between $\vp$ and SM
particles will be maintained as long as $\vp$ and the SM particles remain  relativistic.
At low energies we show that we can re-generate the dark matter field $\phi$ at $E_{gen} $
using the same interaction term $\lin$ as for the high energy inflaton $\phi$ decay.
The appearance of the dark matter field $\phi$ at late times
is then via a quantum transition  and not due to a classical evolution.
A requirement on the mass $\mmp\equiv V''(\phi)$ of the inflaton-dark matter field $\phi$ is that
$ \mp(E_I)> E_{Dinf}\gg E_{gen} > \mp(E_{gen})$ so the simplest potential
$V=\mpo^2\phi^2$ will not work.

The unification scheme presented here has three parameters, the
mass of the dark matter particle $m_o$, the parameter in the inflation
potential $\lm$ and the coupling $g$ for the inflaton decay.
Density perturbations normalized to COBE fixes the value for $\lm$
and the correct amount of dark matter
determines the cross section of dark matter at decoupling (wimp particle)
gives a constraint between $g$ and $m_o$, leaving
only the mass of the dark matter particle $m_o$ as a free parameter. We will show that the
same coupling strength that gives the inflaton decay gives
the dark matter re-generation at low scales and sets in combination with
$m_o$ the wimp decoupling cross section.
These same three parameters are present in models
with inflation and a dark matter wimp particle \ci{wimp} but without unification. This implies
that our unification scheme does not increase the number
of parameters and it accomplishes the desired unification between
inflaton and dark matter for free.

\section{General Framework}

Our starting point is a flat FRW universe with the inflaton-dark
matter field $\phi$ coupled to a relativistic scalar $\vp$. This
field $\vp$ can be either a fermion field or scalar field and we
only require that it is relativistic at least for energies larger than
the mass of the dark matter particle $m_o$. For presentation
purposes we take $\vp$ as a scalar field, but  generalizing
this work to a fermion field is straightforward and the
fermion field  could well be part of the standard model SM
as for example a neutrino. We take the
lagrangian $L=L_{SM}+\widetilde{L}$, where $L_{SM}$ is the
standard model SM lagrangian and
\be
\widetilde{L}=\fr{1}{2}\pu
\phi \pU \phi + \fr{1}{2}\pu \vp \pU \vp -V(\phi)-B(\vp)+
\lin(\phi,\vp,SM),
\ee
$V(\phi), B(\vp)$ are the scalar
potentials for $\phi,\vp$ and $\vin=-\lin$  is the interaction
potential. The classical evolution of $\phi$ and $\vp$ are given
by the equations of motion, \bea
\ddot\phi+3H\dot\phi+V'+ \vin' &=& 0\\
\ddot\vp+3H\dot\vp+B_\vp+ \vip &=& 0
\eea
with $H^2\equiv(\dot a/a)^2= \rho/3$, a prime denotes derivative w.r.t. $\phi$,
$B_\vp\equiv \pp B/\pp \vp$ and we take natural units $m_{pl}^2=1/8\pi G\equiv 1$.
The mass of $\phi$ is given by
\be
\mp^2(\phi)\equiv V''(\phi)
\ee
and it is in  general a field and time dependent  quantity.

\subsection{Inflaton-Dark Matter  Potential }

\subsubsection{Inflaton Potential }

There are many potentials $V(\phi)$ that lead to an inflation
epoch. Inflation with a single scalar field can be
classified in small or large  field models.
Small fields are potentials that inflate for values of the inflaton
$\phi\ll m_{pl}$ as in new inflation models, e.g. $V=V_o(\phi^2-\mu^2)^2$,
while large field models are when inflation occurs for $\phi> m_{pl}$ as
in chaotic models, e.g. $V=V_i\phi^n$ .
Here we will assume the   simplest
inflaton potential because we are more interesting in showing how
the inflaton-dark matter unification scheme takes place in the context of
quantum re-generation process. However, it is  simple to
work with other inflaton potentials.

The potential $V(\phi)$  must satisfy at the inflation scale the slows
roll conditions $|V'/V|<1, |V''/V|<1$ and the constraint on the energy
density perturbation normalized to COBE \ci{wmap}
\be\la{dr}
 \fr{\delta \rho}{\rho}= \fr{1}{ \sqrt{75\pi^2}}\fr{V^{3/2}}{V'}=1.9\times 10^{-5}.
\ee
For example  for chaotic potentials  $V_2\equiv m_o^2\phi^2/2$ or
$V_4\equiv \lm \phi^4$ inflation occurs for $\phi>\phi_e$
and the value of the mass  $m_o$ and the dimensionless parameter $\lm$
and the scale of inflation $E_I\equiv V(\phi_e)^{1/4}$ are given by
\bea\la{v2}
V_2 \equiv \fr{1}{2}m_o^2\phi^2, &&\;  |\phi_e| = 2m_{pl},\\
 m_o = 8.7\times 10^{14}GeV, &&\; E_I=5\times 10^{16} GeV
 \eea
and
 \bea
V_4\equiv \lm \phi^4,&&\;      |\phi_e| = 4m_{pl},\\
 \lm= 10^{-9},&&\, E_I=5\times 10^{16} GeV.
\la{v4}\eea

\subsubsection{Dark Matter Potential }

To obtain Dark Matter at low energies
the scalar potential $V(\phi)$ must have  the limit
\be
V(\phi) \rightarrow V_2 \equiv \mmpo \phi^2
\ee
with $\mpo$ a constant mass term.  A scalar field with potential
$V(\phi)=V_i\phi^{n}$ redshifts as $a^{-3(1+w)}$ with $w=(n-2)/(n+2)$ (for n-even) and only for
$n=2$ do we have a energy density redshifting as matter, $V_2\propto \phi^2 \propto a^{-3}$
while for $V_4\propto \phi^4 \propto a^{-4}$\ci{mio.Q}. The constraint on $V_2$
is that
\be
\rp(t_o)=\fr{1}{2}\dot\phi_o^2+V_2(t_o)=2V_2(t_o)=m_o^2\phi^2_o=\rdmo
\ee
where we used that the pressure vanishes $p= \dot\phi^2/2- V_2=0$ and
$\rdmo$ is the present time dark matter density (from
here on the subscript $o$ gives present time quantities).

\subsection{ Interaction Term}

The interaction term $\lin(\phi,\vp)$ between $\phi$ and $\vp$ should  allow the following processes:\\
i) at high energies $E_{Dinf}$ the inflaton $\phi$ must decay into $\vp$  efficiently, \\
ii) at much lower energies $E_{gen}$  it should re-generate $\phi$ (with $E_{gen}>m_o$),\\
iii)  it must inhibit the decay of dark matter particle $\phi$.\\
The point (iii) is needed because dark matter must last for a long
period of time, however, depending on the ratio of its decay it may be
relax and we could have for example a dark matter dark energy interaction.
However, we will not follow this interesting path here and  will be
presented elsewhere \ci{dm-de.fabio}.
There are many possible interaction terms between $\phi$ and $\vp$ and
generalization of our work presented here is straight forward. If
we take only renormalizable terms with dimensionless constant
we have $\lin=g\phi^a\vp^b$ with $a+b=4$.
We  show in the Appendix that the terms $L_{13}=g\phi \vp^3$
and $L_{31}=g\phi^3 \vp$, i.e. $a=1,b=3$ and $a=3,b=1$ do  not satisfy the conditions
(i-iii). The term $L_{31}=g\phi^3\vp $ does not allow for $\phi$ to decay into $\vp$
if  $\mp\neq m_\vp$ and our working hypothesis has a massive inflaton
field $\phi$ and a relativistic field $\vp$, i.e. $\mp\neq m_\vp$.
On the other hand the term $L_{13}=g\,\phi \,\vp^3 $ does not
work because it gives a decaying dark matter particle. Therefore,
 we are left  with the interaction term
\be\la{l22}
L_{22}(\phi,\vp)= g\,\phi^2 \,\vp^2.
\ee
To produce all SM particle the field $\vp$ must be either
part of the SM or it must couple to at least one field of the SM.
It is standard to assume the interaction term as \ci{inflationarycosmology}
\be\la{vi2}
\lin(\vp,SM)=h\,\vp^2\chi^2+ \sqrt{h}\,\vp \,\bar\psi\psi
\ee
where  $\chi,\psi$  are SM particles.
We will choose $h$ such that the transition $\vp \leftrightarrow SM$
takes place as soon as the $\vp$ are produced.
If the  fields $\chi,\psi$  acquire a large mass
then they will decouple from $\vp$  at $T< m_\chi, m_\psi$ since below
this  temperature $n_\chi, n_\psi$ are exponentially suppressed
and $\G/H$ will be smaller than one. However, as long as $\vp$ remains
relativistic the $\vp$ temperature  redshifts as $T\propto 1/a(t)$
and since it was in thermal equilibrium with the  SM we can
determine $T_\vp(t)=q T_\gamma(t)$ with $q=(g_{sm}(t)/g_{dec})^{3/4}$ and
$g_{dec}$ the number of relativistic degrees of freedom of the
SM at   decoupling.

\subsubsection{Interaction rates }

The   differential transition rate is given by \ci{G}
\be\la{dg}
d\G=V_l(2\pi)^4
|M_{ab}|^2 \delta^4 (P_I-P_F)\Pi_a \fr{1}{2E_a V_l \Pi_b}
\fr{d^3p_b}{2E_b (2\pi)^3}
\ee
 where $P_I(P_F)$ is the initial (final)
momentum, $V_l$ is the volume (normalized to one particle
per volume) and  $M_{ab}\equiv \langle b|M|a\rangle$ is the
transition amplitude. The conservation of energy-momentum requires
that initial and final energies are equal, $E_i=E_f$ and
$p_i=p_f$, and this ensures that the amount of homogeneity of the universe
is preserved by the interaction.
In a process of $a$   initial particles with the same
energy $E_a$  and a final state consisting of $b$
particles with the same energy $E_b$ and   one has
$E_i=a\,E_a =bE_b=E_f$ and the differential transition rate is
given by
\be\la{gab}
 \G_{ab}=c_{ab} |M_{ab}|^2n_a^{a-1}p_b^{b-1}E_a^{b-a-3}
 \ee
with $c_{ab}=(2\pi)^{3-2b}2^{-(a+b)} (a/b)^{2(b-2)}$.

\subsubsection{Interaction rates for $\lin=g\phi^2\vp^2$}

The decay rate for $\phi \rightarrow \vp+\vp $
using eq.(\ref{gab}) with $a=1,b=2$ is given by
\be\la{g12}
\G_{12}=\fr{c_{12} |M_{12}|^2}{ \mp(\phi)}=\fr{c_{12} g^2\phi^2}{\mp(\phi)}
\ee
with $c_{12}=1/16\pi$ and  the interaction term in eq.(\ref{l22})
and we used in the last equality in eq.(\ref{g12}).
The $2 \leftrightarrow 2$ process is given by
\be\la{g22}
\G_{22}=\fr{\widetilde{c}_{22}|M_{ab}|^2 n_a }{ E_a^{2}}=\langle \sigma v \rangle n_a.
\ee
with $\widetilde{c}_{22}=1/32\pi^2$
and where $\langle \sigma v \rangle$ is the cross section times the relative velocity
$v$  of the initial particles  with
\be\la{ss}
\langle \sigma v \rangle=\fr{\widetilde{c}_{22}|M_{ab}|^2 }{ E_a^{2}}=\fr{g^2}{32\pi^2E_a^2}
\ee
and  the in last equality in eq.(\ref{ss}) we used eq.({\ref{l22}).
If one of the initial particles becomes non-relativistic then $n_a$
in eq.(\ref{g22}) is exponentially suppressed by $m/T$
with $n_a=g_a(mT/2\pi)^{3/2}exp[-(m-\mu)/T]$.
However,
if the two initial particles are relativistic than the
number density is given by $n_a =g_a \zeta[3]\pi^2T^3/30 = c_nE ^3 $ with  $E= \bar{r} T$
and $c_n=g_a\zeta(3)/\pi^2\bar{r}^3,\;\bar{r}\equiv (\rho/nT)=\pi^4/30\zeta(3)\simeq 2.7$
and eq.(\ref{g22}) becomes
\be\la{g22r}
\G_{22}=c_{22}\; g^2 E_a
\ee
with $c_{22}=\widetilde{c}_{22}c_n=\zeta(3)/(32 \pi^3 \bar{r}^3)$, and $g_a=1$ for a
real scalar field.

\subsubsection{Decay Rate for  $g \,\phi^2 \vp^2 $}\la{sec22}

The decay process takes place as long as $\mp \gg m_\vp$ and the decay rate $\G_{12}$
for  $\phi \rightarrow \vp+\vp$ is given by
\be\la{d22}
\G_{12}=\fr{ g^2 \phi^2}{\mp(\phi)}
\ee
with $c_{12}=1/16\pi$ and it is field and time dependent trough the terms $\phi,\mp(\phi)$. The field
 $\phi$ in eq.(\ref{d22}) may evolve with time or
 it may be constant if the scalar potential $V(\phi)$
 gives a nonvanishing  v.e.v. $\langle\phi \rangle\neq 0$,
 as in new inflation models.
In order to have a $\phi$ decay we require  that
\be\la{c12}
\fr{\G_{12}}{H}= \fr{ c_{12} \sqrt{3}  g^2 \; \phi^2}{\mp(\phi)\sqrt{V} }
\ee
is larger than one, $ \G_{12}/{H}>1$.  If $\langle\phi \rangle\neq 0$ and $\mp$
is constant then the decay in eq.(\ref{c13}) will be efficient since
$\G$ is constant and $H\propto 1/t$. If however,  the scalar potential
is $V=m_o\phi^2$, as for dark matter, than the decay is not efficient
since $\sqrt{2V}= m_o \phi  \propto 1/t$ and
$\G_{12}/{H}\propto \phi^2/(m_o\sqrt{V})=\phi/m_o^2\propto 1/\sqrt{t}\rightarrow 0$.
In general, the evolution of $\phi$, $\mp$ and  $\G_{12}/{H}$ will depend on
the choice of  scalar potential $V$.

\subsection{ Decay Efficiency}

The interaction or decay process depends on the transition
 rate $\G$ and $H$ and it takes place if the ratio $\G/H>1$.
 The functional form of $\G$ depends on the interaction term $L_{int}(\phi,\vp)$
 and it may be field and time dependent. The classical evolution of the fields depends
 on the scalar potentials $V(\phi), B(\vp)$ and it is then the combination
 of $L_{int}(\phi,\vp)$ and $V(\phi), B(\vp)$ which  determines the
 transition process.

The inflaton decay can be efficient or not efficient.
The process is  not efficient when  $\G(t)/H(t) \leq  1 $ for
$t>t_{dec}$, where $t_{dec}$ is the decoupling time when $\G(t_d)/H(t_d)=1$.
In this case we will have a remanent energy density $\rp(t_d)$.
The energy density $\rp(t)$ will evolve classically depending on the form of the potential
$V(\phi)$. If the potential evolves as matter  then the amount of energy density $\rpd$
at decoupling  is easily determined and it is given by
\bea
\fr{\Omega_{dm i}}{\Omega_{r i}}= \fr{\Omega_\phi(E_d)}{\Omega_r(E_d)}
&=&\fr{\rp}{\rho_r}= \fr{\rpo}{\rho_{ro}}\le(\fr{a_d}{a_o}\ri)=
\fr{\Ompo}{\Omega_{ro}} \le(\fr{E_o}{E_d}\ri) \nonumber  \\
&=&  3\times 10^{-23} \le(\fr{10^{14} GeV}{E_d}\ri)
\la{opi}\eea
with $\Opo=\Odmo=0.22$ and $ \Omega_{ro}h^2= 4.15\times 10^{-5}$ the present time relativistic energy density.
Clearly one requires a huge amount of fine tuning in the value of $\rp$ at
a high decoupling energy since our universe had a large radiation domination epoch.
We conclude that if we only take into account a classical evolution of $\phi$
after reheating the inflaton-dark matter unification requires a large fine tuning
of initial conditions.

On the other hand  the transition process is efficient when the inflaton decays completely and this requires that
 $\G(t)/H(t) > 1 $ for $t>t_d$, where $t_d$
is the time when $\G(t_d)/H(t_d)=1 $.  In this case
the $\phi$ particles decay completely  and disappear from the spectrum
and $\Op\rightarrow 0$. A simple example is when
$\G$ is constant since $H\propto 1/t$ and  $\G/H\propto t \rightarrow \infty$.
However if $\G(t)/H(t)$ becomes smaller than one for $t>t_{dec}$ then
we will also say that the decay is efficient if the residual energy density $\rp(t)$
is subdominant. For example if $\rp(t)$ redshifts as matter then   that
the decay is efficient if $\rp(t)\ll  \rdm(t)$ for $t>t_{dec}$
with $\rdm=\rdmo (a_o/a)^3$ the dark matter energy density.
An efficient decay would clearly not allow  $\rp$  to account for
dark matter.

\section{Inflaton-Dark Matter Unification with Quantum Generation}

\subsection{Generic Quantum Transitions }

Another possibility to achieve inflaton-dark matter unification
is if the $\phi$ particles  are re-generated via a quantum
transition \ci{fabios} at some late time but before $a_{eq}$, the
matter-radiation equivalence scale factor. In this case, the decay
$\phi\rightarrow \vp$ at high energies $E_{Dinf}$, i.e. below inflation, must be
efficient and $\phi$ disappears from the spectrum.
At much lower scales $E_{gen}$, with $E_{gen}\ll E_{Dinf}$, the
$\phi$ can be re-generated by $\vp$.

After inflation the energy of the $\phi$ particles is $E_I^2=p_I^2+\mp(E_I)^2\simeq \mp(E_I)^2$
since the momentum $p_I$ redshifted with the expansion of the universe, i.e. $p_I=e^{-N}p_{I i}$
with $N$ the number of e-folds of inflation.
If $\phi$ decays into $\vp$ at $E_{Dinf}$ we will produce relativistic $\vp$ particles with energy
$ E_{Dinf} < \mp(E_I)$.
On the other hand at low energies $E_{gen}$ when we have the inverse process of $\phi$
production from $\vp$ we must have that the energy of the $\vp$ particles
$E_{gen}$ must satisfy  $E_{gen}> \mp(E_{gen})$ and therefore we must have
\be\la{mmm}
\mp(E_I)> E_{Dinf}\gg E_{gen} > \mp(E_{gen}),
\ee
which implies that the mass of $\phi$ after inflation must be much larger than at
generation scale  $E_{gen}$. Since $\mp(E_d)\equiv V''(E_d) \gg  \mp(E_{gen})\equiv V''(E_{gen})$
clearly the simplest inflationary potential $V=m_o\phi^2/2$, with $m_o$ constant,
will not work. However, potentials such as $V=m_o\phi^2/2+\lm \phi^4$
or of the new inflation type, e.g. $V=V_o(\phi^2-\mu^2)^2$, where the $\phi$ field rolls down
a flat region for small $\phi$ and then oscillates around the  minimum of the potential at $\langle\phi\rangle=\mu$
may work in this scenario.

\section{Inflaton-Dark Matter Unification Model}

\begin{figure}[tbp]
\begin{center}
\includegraphics*[width=8cm]{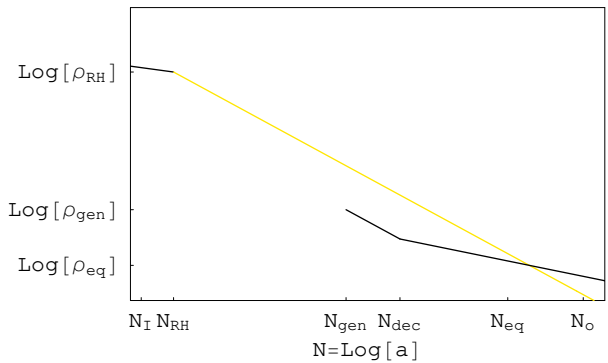}
\caption{We show the evolution of the inflaton-dark matter field
$phi$ (black) and of the relativistic (yellow) energy densities.
Notice that between $N_{RH}$ and $N_{gen}$ there is no
$\rp$ (no $\phi$ particles) and that the evolution of $\rp$ goes from relativistic
to dark matter type at $N_{dec}$.}
\label{fig0}
\end{center}
\end{figure}
We will work out the inflation-dark matter unification through a simple
example. Of course, the whole scheme is much more general and
other inflaton-dark matter $V(\phi)$ potentials or interaction terms may be used,
however in all cases $V(\phi)=m_o^2\phi^2$ for dark matter.
We take the following potential
\be\la{v}
V(\phi)= V_2+V_4=  \fr{1}{2}m_o^2\phi^2+\lm \phi^4
\ee
with a  mass
\be
\mp^2 =  m_o^2 + 12\lm\phi^2
\ee
and we coupled $\phi$ with $\vp$ via the interaction term
$L_{22}=g\,\phi^2 \,\vp^2 $. For
\be
\phi\geq  \phi_{24}\equiv \fr{m_o}{\sqrt{2\lm} }
\ee
we have $V_4\geq V_2$. The energy at $V_2=V_4=V/2$ is
\be
E_{24}\equiv V(\phi_{24}) = 2 \le(\fr{m_o^2 \phi_{24}^2}{2}\ri)^{1/4}= \le(\fr{4}{\lm}\ri)^{1/4}m
\ee
So, if $\lm\neq 0, m_o\neq 0$
the potential  $V_4$ dominates at high energies, during inflation,
while $V_2$ at low energies, when dark matter prevails. If either  potential
$V_4$ or $ V_2$ dominates then its classical evolution is  $V_4\propto \phi^4 1/a^4$ with $\phi  \propto 1/a$
and $V_2 \propto \phi^2 1/a^3$ with $\phi  \propto 1/a^{3/2}$. From eq.(\ref{v4}) we know
that inflation requires $\lm=10^{-9}$ and inflation ends at $\phi=2.6$ an energy
$ E_I\equiv V_4^{1/4}(\phi_e)=5\times 10^{16}\, GeV $ and a mass $\mp=V_4''=\sqrt{12\lm}\phi$
and $\mp(E_I)\simeq 10^{14}\,GeV$. At low energies the mass of the dark matter
particle is given by $m_o$ and for CDM the mass of $\phi$ must be $m_o\geq O(GeV)$ while warm DM requires
a smaller mass with $m_o> O(10-100)\, keV$ \ci{warm}.

We will use in this model the interaction term
\be\la{22}
L_{22}= g\phi^2\vp^2.
\ee
This term will allow the inflaton to decay efficiently  after inflation  at high energies $E_{Dinf}$
into the relativistic $\vp$ particles and $\phi$ will disappear from
the spectrum (at most $\Op$ is negligible). At a much later time the same interaction term
in eq.(\ref{22}) will produce relativistic particles $\phi$ at energies below
$E_{gen}$ with $E_{gen} \gg m_o$.  Eventually the $\phi$ particles become non-relativistic
and they will decouple from $\vp$ as a WIMP particle. The constraint to
give the correct amount of dark matter today fixes the cross section $\sigma = g^2/(32\pi m_o^2)$.
Finally, since $\phi$ becomes massive we also show that $\phi$ at low
energies    does not   decay
into $\vp$ and allows for dark matter to dominate  a long period as in a standard cosmological scenario.
We show in fig.\ref{fig0} the evolution of the
inflaton-dark matter and relativistic energy densities. We see that for a
long period of time the field $\phi$ disappears from the spectrum
from the inflaton decay to the re-generation scale.  Since
the $\phi$ is re-generated while relativistic $\rp$ redshifts as
radiation first and then when it becomes non-relativistic it redshifts
as matter and $\phi$ decouples from $\vp$ as any WIMP particle.

\subsection{Quantum Transitions}

The inflaton-dark matter field $\phi$ should decay efficiently
at high energies, after inflation,  to reheat the universe but at low energies
when dark matter dominates we do no longer want the $\phi$ to decay
since the period of dark matter domination must last from
$a_{eq}$ to $a_{de}\simeq  a_o/2$ when dark energy starts to dominate.

\subsubsection{$\phi$ Decay at Inflation: $E_{Dinf}$}

The interaction term in eq.(\ref{22}) gives a decay
$\phi \rightarrow \vp +\vp$ with a decay rate given by
eq.(\ref{g12}) and at high energies
when $V_4$ dominates $V$ one has $\mmp=12\lm\phi^2$ and
\be\la{gdi}
\G_{Dinf}= c_{12 }g^2\fr{\phi^2}{\mp}=c_{12}g^2 \fr{\phi}{\sqrt{12\lm}}
\ee
with $c_{12} = 1/16\pi$. Using  $H=\sqrt{V_4/3}=\phi^2\sqrt{\lm/3}$ we have
\be\la{ghdi}
\fr{\G_{Dinf}}{H}= \fr{c_{12}g^2}{2\lm\,\phi} \equiv \fr{E_{Dinf}}{E_4}
\ee
where $E_4\equiv V_4^{1/4}=\lm^{1/4}\phi$ is an energy scale which
depends on the field $\phi$  and
\be\la{edi}
E_{Dinf}\equiv \fr{c_{12}g^2\lm^{-3/4}}{2}
\ee
is a constant quantity with energy dimensions
and set the scale of the decay. We have  a decay for energies
$E_4<E_{Dinf}$. The evolution of  $E_4 \propto \phi \propto 1/a$
and  $\G_{Dinf}/H\propto 1/\phi \propto a$ grows with time giving  an efficient decay. Once $\G_{Dinf}/H>1$
the decay of $\phi$  takes place and does not stop (as long as $V_4$ dominates $V$).

\subsubsection{Quantum Re-generation: $E_{gen}$}

If the field $\phi$ decays completely after inflation
than there will no $\rp$ left to account for dark matter.
In order to re-generate the field $\phi$ we follow \ci{fabios}
and we use the same interaction term $L_{22}$ as for the inflaton decay
but now  the universe contains $\vp$  relativistic particles and no $\phi$ particles.
As long as the energy $E_\vp$ of the relativistic particles
is larger than the mass of $\phi$, i.e. $E_\vp> m_o$, we can
produce $\phi$ particles. If $E_\vp \gg m_o$
then both fields $\vp,\phi$ are relativistic and the transition rate
for the $2\leftrightarrow 2$  process is given by eq.(\ref{g22r}) with
\bea\la{dbd}
\G_{gen}=c_{22} g^2 E ,&&\;\; H=\sqrt{\fr{\rho_r}{3\Omega_r}} =c_H E^2,\\
\fr{\G_{gen}}{H}&&=\fr{c_{gen}\, g^2  }{E}\equiv \fr{E_{gen}}{E}
\eea
with $c_{22}=\zeta(3)/(32 \pi^3 \bar{r}^3)$, $ c^2_H\equiv g_{rel}\pi^2/(90\Omega_{rel}\bar{r}^4)$
and $c_{gen}\equiv  c_{22}/c_H= (\zeta(3)/32 \pi^4 \bar{r}) \,
(90 \, \Omega_{rel}/g_{rel})^{1/2}\simeq 10^{-4}$ with
$g_{rel}\simeq 106,\Omega_{rel}\simeq 1$. We have taken in eq.(\ref{dbd})
that $n_\phi$ is proportional to $T^3$ since $\phi$ is relativistic.
The process takes place for energies $E\propto T\propto 1/a$ of the relativistic
particles below the constant scale $E_{gen}$ with
$\G_{gen}/H>1$ or
\be
E\leq E_{gen}\equiv c_{gen} g^2.
\ee
Since the particles $\vp$ are relativistic
and they are in thermal equilibrium with  SM particles we have $T_\phi=T_\vp\simeq T_{sm} $.
As long as $T\gg m_o$ we will $\Omega_\vp=\Omega_\phi$, but once we reach
the region with $T\gtrsim m_o$ the two fields will decouple
since $n_\phi$ will be exponentially suppressed.

\subsubsection{Non-relativistic $\phi$ Decoupling: $E_{dec}$}\la{secdec}

If two relativistic particles are in thermal equilibrium and one
(in our case $\phi$) becomes non-relativistic
then the density number $n_\phi$ is exponentially suppressed and
$\vp$ and $\phi$ decouple. This is just the standard WIMP particle
decoupling. The
transition rate  for a $2 \leftrightarrow 2$ process is given by
eq.(\ref{g22})
\be\la{gdec}
\G_{dec}=\langle \sigma v \rangle n_\phi.
\ee
In order to
have the correct amount of dark matter a WIMP must decouple at \ci{wimp}
\be\la{os}
\Omega_\phi h^2=\fr{3\times 10^{-27}cm^3 s^{-1} }{\langle \sigma v \rangle}
\ee
For the $2\leftrightarrow 2$ transition this implies a cross section
\be\la{s}
\langle \sigma  \rangle = \fr{g^2}{32\pi m_o^2}=0.1 pb
\ee
with  $v\simeq c$.
Eq.(\ref{s}) gives a constraint for $g$ in terms of the mass $m_o$. The freeze out
takes place at $x_F=m_o/T_F\simeq 10$ \ci{wimp}  giving a decoupling constant energy
$E_{dec}$ which is a function of  the dark matter  mass
\be\la{edec}
E_{dec}= c_{dec} T_{F}= c_{dec} \fr{m_o}{x_F}\simeq 0.12\; m_o
\ee
with $c_{dec}=(\pi^2 g_{rel}/30)^{1/4}\simeq 1.2$ with $g_{rel}\simeq 5.5+1=6.6$
at $E<O(MeV)$.
For energies $E<E_{dec}$ the fields $\phi$ and $\vp$ are no longer coupled
and $\phi$ evolves classically as matter with $V\simeq V_2\propto \phi^2 \propto 1/a^3$.
If $m_o\gg T_{eq}\simeq eV$, where $T_{eq}$ is the radiation-matter equality,
the decoupling takes place while the universe is radiation dominated
and the constraint on warm dark matter sets a lower scale $m_o> 10-100 keV$ \ci{warm}.

\subsubsection{Dark Matter Decay?: $E_{Ddm}$}

We have seen that at $E_{dec}$ the field $\phi$ ceases to maintain thermal equilibrium
with $\vp$ through the $2\leftrightarrow 2$ process. However,
the field $\phi$ may decay into $\vp$ since $\mp\gg m_\vp$.
Of course, we do not want $\phi$ to decay since it must account
for dark matter. In this case the decay rate is the same is in eq.(\ref{g12})
\be\la{ddm}
\G_{Ddm}= \fr{c_{12}g^2\phi^2}{m_o}
\ee
but the mass $m_o$ is now constant and $\phi$  evolves as $\phi\propto a^{-3/2}$ since
now the potential that dominates is $V_2\gg V_4$ and $\rp=2V_2=m_o^2\phi^2\propto 1/a^3$. For
radiation dominated epoch we have $ H=\sqrt{\fr{\rho_r}{3\Omega_r}} \equiv c_H E^2$
and
\be
\fr{\G_{Ddm}}{H}=\fr{c_{Ddm}g^2\phi^2}{m_o E^2}
\ee
with $c_{Ddm}\equiv c_{12}/c_H\simeq 10^{-3}$. Since
$E=E_o(a_o/a)$, $\phi=\phi_o(a_o/a)^{3/2}$ then $\phi^2/E^2=(\phi_o^2/E_o^2)(a_o/a)=(\phi_o^2/E_o^2)(E/E_o)$
 we can write
\be\la{ghdm}
\fr{\G_{Ddm}}{H}=\fr{c_{Ddm}g^2\phi_o^2\,E}{m_o E_o^3} \equiv \fr{E}{E_{Ddm}}.
\ee
We will not have a $\phi$ decay  for  $\G_{Ddm}/H<1$, i.e. energies $E$ below the
constant energy $E_{Ddm}$,
\be\la{edm}
E< E_{Ddm} \equiv \fr{m_o \;E_o^3}{c_{Ddm} g^2\phi_o^2}\simeq 10\; m_o
\ee
where we have used   $\phi_o = \sqrt{\rdmo}/m_o$, $E_o=  \bar{r}T_o$
and eq.(\ref{s}). From eq.(\ref{edec}) and (\ref{edm})
we have
\be
\fr{E_{Ddm}}{E_{dec}}\simeq \fr{10\; x_F}{c_{dec}}\simeq 82>1
\ee
so that after decoupling at $E_{dec}$ there is no $\phi$ decay.

In the region when dark matter dominates the universe we can easily estimate
$ \G_{Ddm} /H$ as
\bea\la{ghdm2}
\fr{\G_{Ddm}(t_r)}{H(t_r)}\fr{H(t_m)}{\G_{Ddm}(t_m)} =&&
\fr{\phi^2(t_r)}{\phi^2(t_m)}\fr{H(t_m)}{H(t_r)}\nonumber\\
=&&\fr{a(t_m)}{a(t_r)}\sqrt{\fr{a(t_m)}{a(t_{eq})}}>1
\eea
since $a(t_r)<a(t_{eq})<a(t_m) $ where $t_r,t_{eq},t_m$ are times in
radiation, equality and matter domination and $H(t_r)\propto
a^{-2},H(t_m) \propto a^{-3/2}$. From eq.(\ref{ghdm}) and (\ref{ghdm2})
we have $1> \G_{Ddm}(t_r)/H(t_r)>\G_{Ddm}(t_m)/H(t_m)$ for
energies $E<E_{Ddm}$,  so we conclude that neither in radiation
nor in matter domination can $\phi$ decay into $\vp$.

\subsection{ Universe Reheating}

The reheating of the universe takes place via a process
$\vp+\vp\leftrightarrow \chi+\chi$ (or
$\vp+\vp\leftrightarrow\psi+\psi$)  with a cross section for
relativistic particles $\sigma = h^2/32\pi E^2$ (we take the same
strength for   the $\chi $ and $\psi$) and an interaction rate
\bea\la{gdr}
 \G_{R}=   c_{22} h^2 E
,&&\;\;\;\; H=\sqrt{\fr{\rho_r}{3\Omega_r}} \equiv c_H E^2,\\
\fr{\G_{R}}{H}=\fr{c_R\,  h^2  }{E}\equiv \fr{E_R }{E} ,&& \;\;\;\;E_R\equiv c_R h^2
\eea
with $c_{22}=\zeta(3)/(32 \pi^3 \bar{r}^3)$, $ c^2_H\equiv g_r\pi^2/(90\Omega_r\bar{r}^4)$  and
$c_R\equiv c_{22}/c_H $.
For $E> 10^2 GeV$ we have $g_r\simeq 106, \Omega_r\simeq 1$ and
$c_R\simeq 10^{-3}$. Clearly eq.(\ref{gdr})
maintains a TE for $E\leq E_R$. A good choice of $h$ is
such that reheating takes  as soon as the $\vp$ particles
are produced at $E_{Dinf}$ with at  $E_R=c_H h^2= E_{Dinf}$
but if can take any values as in the range $E_{Dinf}> E_R>O(10-100)MeV$
which is the lower limit for reheating \ci{lowinflation}
\begin{figure}[tbp]
\begin{center}
\includegraphics*[width=8cm]{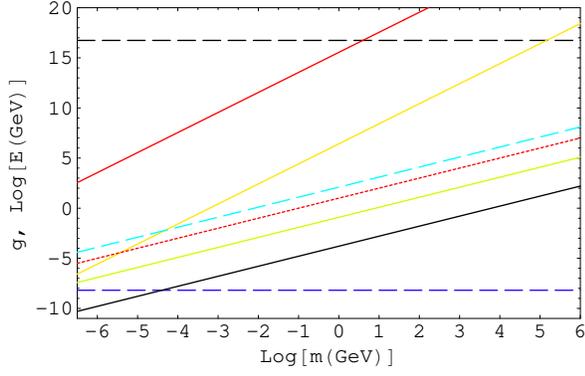}
\caption{We show the dependence of the different energies
densities as a function of $m_o$. The dashed lines
give the scales of $E_I,E_{24},E_{eq}$ (black, light blue,
blue, respectively), while the solid lines are from top
to bottom $E_{Dinf},E_{gen},E_{Ddm},E_{dec}$ all in GeV
(red, yellow, dotted-red, light green, respectively).
The black solid line is the dimensionless coupling $g$.
Notice that $E_I>E_{Dinf}>E_{gen}> E_{24}>E_{Ddm}>E_{dec}>E_{eq}$ }
\label{fig1}
\end{center}
\end{figure}

\subsubsection{Summary of Energies}

We present the different energy scales relevant in the
process of our inflation-dark matter unification scheme.
From high energy to low energy we have the following energy scales:
Inflation occurs at $E_I$, then $\phi$ decays efficiently into  $\vp$ at $E_{Dinf}$
via the interaction term $L_{22}$ and disappears from the  spectrum.
Using the same interaction term
the field  $\phi$  is re-generated at a much lower scale $E_{gen}$.
The field $\phi$  becomes none relativistic while in thermal
equilibrium with $\vp$ and decouples at $E_{dec}$.
We also show that below $E_{Ddm}$ the field $\phi$ does not decay again into $\vp$
with  $E_{Ddm}/E_{dec}=8 m_o/T_F>1$, which ensures that after thermal decoupling the field
$\phi$ does not further decay.
We have then the following order of energies
\be\la{ee}
E_I>E_{Dinf}>E_{gen}> E_{24}>E_{Ddm}>E_{dec}>E_{eq}.
\ee
SM particles are produced at $E_R$ via the coupling with $\vp$ and the constraint
is  $ E_{Dinf}>E_{R}>O(10-100)MeV$ \ci{lowinflation}.

Concerning the inflaton-dark matter unification scheme
we have 3 different parameters $\lm, m_o$ in the potential $V$
and a coupling $g$ between  $\vp$ and $\phi$. The seven energies
in eqs.(\ref{ei})-(\ref{ef})  are given in terms of these three parameters.
Inflation fixes one parameter,  $\lm$, and the   amount
of dark matter today gives a constraint between
$g$ and $m_o$.  We are left with one single free parameter
which we take it to be the mass $m_o$.  We show
in fig.(\ref{fig1}) the dependence of the $E's$ on $m_o$
and in table 1 we give the values for $10^{-4}GeV<m_o< 10^{4}GeV$.
We see that $E_{dec}>E_{eq}=O(10^{-9}GeV)$ and that the values of $g$
and all other energies are phenomenologically viable. This
implies that it is feasible to implement the inflation-dark matter
unification. We would like to point out
when $E_{Dinf}> E_I$, or even larger than $m_{pl}$ this only means that
as soon as $\phi$ ends inflation at $E_I$,  $\phi$ decays immediately  since
the condition for its decay is satisfied.
We resume the definitions and values of these energies
\bea\la{gg}
g&=&(32\pi \langle \sigma \rangle )^{1/2}m_o=1.6\times 10^{-4} \le(\fr{m_o}{GeV}\ri)\\
\la{ei}E_{I} &\equiv & \lm^{1/4} \phi_e=5.4\times 10^{16}  GeV \\
E_{Dinf} &\equiv &\fr{c_{12} g^2 \lm^{-3/4}}{2}=\fr{16\pi c_{12} \langle \sigma \rangle \,m_o^2}{\lm^{3/4} }
\\ &=& 3.5\times10^{15} \le(\fr{m_o}{GeV}\ri)^2 GeV\nonumber \\
E_R&\equiv & c_R h^2\\
E_{gen} &\equiv & c_{gen}g^2= 32\pi c_{gen} \langle \sigma \rangle \, m_o^2
\\ &=& 2.6\times 10^{6} \le(\fr{m_o}{GeV}\ri)^2 GeV \nonumber    \\
E_{Ddm} &\equiv & \fr{E_o^3\;m_o }{c_{Ddm} g^2 \phi_o^2 }
= \fr{(rT_o)^3\; m_o}{32\pi\,  c_{Ddm}\,\rdmo \langle \sigma \rangle}
\\ &=& 10 \le(\fr{m_o}{GeV}\ri) GeV  \nonumber \\
E_{dec} &\equiv & c_{dec} T_F = \le(\fr{\pi^2g_{rel}}{30} \ri)^{1/4}
\fr{m_o}{x_F}
\\ &=& 0.1  \le(\fr{10}{x_F}\ri) \le(\fr{m_o }{GeV}\ri) GeV  \nonumber \\
E_{24}&\equiv &\le(\fr{m_o^2 \phi_{24}^2}{2}\ri)^{\fr{1}{4}} = \le(\fr{4}{\lm}\ri)^{1/4}m_o
\\ &=& 125  \le(\fr{m_o}{GeV}\ri) GeV.\nonumber
\la{ef}\eea
\begin{table}
\begin{center}
\begin{tabular}{|c|c|c|c|c|c|c|c| }
  \hline
  m  & g &  $E_I$ & $E_{Dinf}$ & $E_{gen}$ & $E_{Ddm}$ &  $E_{dec}$   & $E_{24}$ \\
  \hline\hline
  $10^{-4}$ & $ 2\cdot 10^{-8} $    & $ 5\cdot   10^{16} $    & $4\cdot  10^{\,7}  $  & $  3\cdot  10^{-2}  $  & $    10^{-3}  $  & $    10^{-5}   $ & $  10^{-2}   $      \\
 $10^{-2}$  & $ 2\cdot 10^{-6} $    & $5\cdot    10^{16} $   & $ 4\cdot 10^{11}  $  & $3\cdot     10^{\,2} $ & $ 10^{-1}  $  & $  10^{-3}    $ &  $ 1 $            \\
  $    1 $ & $ 2\cdot 10^{-4}  $   & $ 5\cdot   10^{16} $    & $4\cdot  10^{15}  $  & $3\cdot     10^{\,6} $  & $    10^{\,\,}   $   & $    10^{-1}   $ & $ 10^{\,2} $        \\
 $10^{\,2} $ &$ 2\cdot  10^{-2}  $    &$  5\cdot  10^{16} $    &$4\cdot 10^{19}   $  & $3\cdot    10^{10}  $  & $    10^{\,3}  $  & $   10^{\,\,}   $ & $ 10^{\,4} $       \\
$ 10^{\,4}  $& $ 2    $     &  $ 5\cdot  10^{16} $   & $  4\cdot 10^{23}  $  & $3\cdot     10^{14} $  & $   10^{\,5}   $  & $   10^{\,3}   $ & $  10^{\,6} $        \\
  \hline
\end{tabular}
\end{center}\la{tabel1}
\caption{\small{We show the values of the energies given in eqs.(\ref{gg})-(\ref{ef}) in $GeV$ as a function
of $m_o(GeV)$ and the value of the dimensionless coupling $g$. We see that $E_{dec}>E_{eq}=0.5\, eV$.}}
\la{constr}\end{table}

\section{Summary and conclusions}

We have presented a model
where inflation and dark matter takes place via a single
scalar field $\phi$. This unification is realized  via
a quantum re-generation of the $\phi$ field at low energies.
The late time  appearance of $\phi$ solves the fine tuning
problem of the initial condition of dark matter  density $\Omega_{dmi}$
at high energies in models without quantum re-generation,
and  allows for having a long lasting radiation dominated universe after
reheating.

The unification scheme presented here has three parameters, the
mass of the dark matter particle $m_o$, the inflation parameter $\lm$ and
the coupling $g$ for the inflaton decay. Phenomenology sets the values
for $\lm$ and gives a constraint  between $g$ and $m_o$, leaving
only the mass of the dark matter particle $m_o$ as a free parameter. We have shown that the
same coupling strength that gives the inflaton decay gives
the dark matter re-generation at low scales and sets in combination with
$m_o$ the wimp decoupling cross section.
These same three parameters are present in models
with inflation and a dark matter wimp particle but without unification. This implies
that our unification scheme does not increase the number
of parameters and it accomplishes the desired unification between
inflaton and dark matter for free.

\noindent{\bf Acknowledgment}
We would like to thank Luis A. Urena for useful discussions
and comments.
We thank for partial support Conacyt Project 80519, IAC-Conacyt
Project.

\appendix

\section{Other Interaction terms  }\la{sec13}

We will nos show that the terms $L_{13}=g\phi\vp^3 $ and $L_{31}=g\phi^3\vp $  do
not allow to implement the inflaton-dark matter unification scheme.
The term $L_{31}=g\phi^3\vp $ does not allow $\phi$ to decay into $\vp$
if  $\mp\neq m_\vp$ and our working hypothesis has a massive inflaton
field $\phi$ and a relativistic field $\vp$, so that $\mp\neq m_\vp$.
This can be easily seen since
the decay process is $\phi \rightarrow \vp$ and   energy momentum conservation
implies that $E_\phi=E_\vp$ and $E_\phi=p_\vp$ which requires the masses
to be equal, $\mmp=E_\phi^2-p_\phi^2=E_\vp^2-p_\vp^2=m^2_\vp$.

On the other hand the term $L_{13}=g\,\phi \,\vp^3 $ will not
work because it gives a decaying dark matter particle.
The decay  process $\phi \rightarrow \vp+\vp+\vp$
has  a decay rate
\be\la{d13}
\G_{13}=c_{13}g^2 \mp(\phi),\hspace{.5cm} H=\sqrt{\fr{\rp}{3}}=\sqrt{\fr{V}{3}}  , \hspace{.5cm}
\ee
with $c_{13}=1/((2\pi)^3 144)\simeq 10^{-5}$  and we have used that the universe is dominated
by $\phi$ and $\rp\simeq V$, valid after inflation and matter domination. We have
\be\la{c13}
\fr{\G_{13}}{H}=c_{13} g^2  \sqrt{\fr{ 3\mmp}{V}}
\ee
and  $\phi$ decays into $\vp$ if $\G/H>1$, i.e. as long as $3c_{13}\, g^4\mmp(t)> V(t)$.
If $\mp$ is constant, as for dark matter, and since $H\propto  \sqrt{V(t)}\propto 1/t \rightarrow 0$
we have $\G/H\propto \mp/\sqrt{V} \propto t \rightarrow \infty$, i.e. we have an
efficient decay.  This is fine  for reheating the universe at high energies but
at low energies, when dark matter dominates, the same process
takes place. This is clearly a problem
since we know that the dark matter potential has a constant mass $m_o$.

If we impose that $\phi$ should not decay until present time
than we must impose that  $\G(t_o)/H(t_o)<1$ for $t>t_o$. Taking
a  dark matter particle  mass $m_o>10^{-18}$, i.e. $m_o>1 GeV$,
and   $V_o=m_o^2\phi^2_o/2\simeq 10^{-120} $ we have
$\sqrt{m_o^2/V(\phi_o)}>10^{42} $. Therefore, $\G/H<1$
gives  a small coupling  $g^2<10^{-42}/c_{13}=10^{-37}$. On the other hand,
for the inflaton $\phi$  to decay with such a small coupling we
need $\G/H>1$ and an inflation potential $V\lesssim  c_{13}^2 g^4 \mmp< 10^{-84}\mmp $
which for $\mp<m_{pl}$ gives an inflation scale $V< MeV^4$.
We can conclude that  the term $L_{13}$
does  not allow for inflation-dark matter unification since it cannot
give a high enough reheating scale without having a low energy
dark matter decay.

\end{document}